\documentclass[5p]{elsarticle}
\usepackage{graphicx,amsmath,amssymb}
\usepackage[breaklinks,colorlinks=true,pdfstartview=FitV]{hyperref}

\newcommand{\rmd}{\mathrm{d}}
\newcommand{\rme}{\mathrm{e}}
\newcommand{\rmi}{\mathrm{i}}
\newcommand{\ti}{t_{\rm I}}
\newcommand{\tf}{t_{\rm F}}
\newcommand{\T}{{\rm T}}
\newcommand{\tT}{\tilde{\rm T}}
\newcommand{\hatrho}{{\hat \rho}}
\newcommand{\hatphi}{{\hat \phi}}
\newcommand{\DF}{D_{\rm F}}
\newcommand{\DR}{D_{\rm R}}
\newcommand{\Dt}{\Delta t}
\newcommand{\calV}{\mathcal{V}}
\newcommand{\calE}{\mathcal{E}}
\newcommand{\bp}{\boldsymbol{p}}
\newcommand{\bA}{\boldsymbol{A}}
\newcommand{\bE}{\boldsymbol{E}}
\newcommand{\teta}{\bar{\eta}}
\newcommand{\tphi}{\bar{\phi}}
\newcommand{\tpsi}{\bar{\psi}}
\newcommand{\hphi}{\hat{\phi}}

\begin{document}

\author[tokyo]{Kenji Fukushima}
\ead{fuku@nt.phys.s.u-tokyo.ac.jp}

\author[tokyo,riken]{Tomoya Hayata}
\ead{hayata@riken.jp}

\address[tokyo]{Department of Physics, The University of Tokyo,
 7-3-1 Hongo, Bunkyo-ku, Tokyo 113-0033, Japan}
\address[riken]{Theoretical Research Division, Nishina Center,
 RIKEN, Wako, Saitama 351-0198, Japan}

\title{%
{\vspace{-2.0cm}
\small\hfill\parbox{3.2cm}{\raggedleft%
RIKEN-QHP-115
}}\\[1.5cm]
Schwinger Mechanism with Stochastic Quantization}

\begin{abstract}
 We prescribe a formulation of the particle production with real-time
 Stochastic Quantization.  To construct the retarded and the
 time-ordered propagators we decompose the stochastic variables into
 positive- and negative-energy parts.  In this way we demonstrate how
 to derive a standard formula for the Schwinger mechanism under
 time-dependent electric fields.  We discuss a mapping to the
 Schwinger-Keldysh formalism and a relation to the classical
 statistical simulation.
\end{abstract}
\maketitle

\section{Introduction}

Direct simulations of the quantum field theory formulated on
discretized space-time, that is, lattice simulations have proved to
be a powerful numerical tool to reveal non-perturbative aspects of the
theory.  It is, however, not always guaranteed that one can dig
meaningful information out from the lattice calculations.  Because the
numerical algorithm relies on the importance sampling, the method
ceases to work as soon as the integrand becomes negative (or complex
in general).  In gauge theories the most notorious example to hinder
the lattice numerical approach is the ``sign problem'' associated with
finite density of fundamental
fermions~\cite{Barbour:1997ej,Alford:1998sd} (for reviews; see
Ref.~\cite{Muroya:2003qs}).  The sign problem is activated also when
the theory has a Chern-Simons term that is necessary to access the
$\theta$-vacuum
structure~\cite{Azcoiti:2002vk,Aoki:2008gv,Vicari:2008jw,D'Elia:2012vv}.

In addition to these Euclidean examples one cannot avoid encountering
the sign problem if one attacks the real-time problem in Minkowskian
space-time.  The complex phase originates from the path-integral
weight, $\rme^{\rmi S}$.  The real-time simulation is one of the most
challenging topics in modern quantum field theories;  the transport
coefficients of a fluid, the particle emission rate in strongly
correlated systems, and so on, are needed in various physics
circumstances.  One can still utilize the conventional lattice
technique as long as the analytical continuation from Euclidean
space-time is a legitimate
procedure~\cite{Asakawa:2000tr,Karsch:2001uw,Nakamura:2004sy,Meyer:2007ic}.
The applicability of such approach is, however, limited to static (or
steady) phenomena or linear-response perturbation at best.  Full
quantum simulations would demand an alternative quantization machinery
in different directions from the importance sampling.  For a promising
candidate, in this work, we will advocate the Stochastic
Quantization~\cite{Parisi:1981,Berges:2005yt} (for reviews, see
Ref.~\cite{stochastic-review}) and take a concrete example of
real-time physics problem.

One of the most important and most ubiquitous phenomena that call for
real-time quantization is the problem of the particle production from
the \textit{vacuum}.  In the quantum field theory, in fact, the vacuum
is not empty but is full of quanta, and some of them could
\textit{tunnel} the potential barrier out from the vacuum.  Celebrated
examples of such tunneling phenomena include the Schwinger mechanism
that refers to the vacuum-insulation breakdown under external electric
fields~\cite{Heisenberg:1935qt,Schwinger:1951nm} (for a review, see
Ref.~\cite{Dunne:2004nc}), and the Hawking radiation that refers to
the spontaneous radiation process from black holes, namely, the
particle production under external gravitational
fields~\cite{Hawking:1974sw,Parikh:1999mf}.

In this work we shall focus specifically on a theoretical
reformulation of the Schwinger mechanism on the basis of the
Stochastic Quantization.  For attempts in different directions the
readers can consult the
literature~\cite{Hebenstreit:2010vz,Semenoff:2011ng,Hashimoto:2013mua}.
Because the Stochastic Quantization is a functional description in
terms of classical fields, we must first establish a prescription to
derive various kinds of propagators which are written most
conveniently with creation/annihilation operators.  In
Refs.~\cite{Fukushima:2009er,Gelis:2013oca,Fukushima:2014sia} it has
been shown that the inclusive spectrum is to be expressed in the
following manner:
\begin{equation}
 \begin{split}
 \frac{\rmd N}{\rmd^3\bp} &= \frac{1}{(2\pi)^3\,2\calE_{\rm out}(\bp)}
  \lim_{t=t'\to\infty} \bigl[ \partial_{t'} + \rmi \calE_{\rm out}(\bp) \bigr] \\
 & \qquad\times \bigl[\partial_{t}-\rmi \calE_{\rm out}(\bp) \bigr]\,
 \bigl\langle \hatrho_{\rm in}\,\hatphi^\dag(t',\bp)\,
  \hatphi(t,\bp) \bigr\rangle \;.
 \end{split}
\label{eq:production}
\end{equation}
The initial density matrix is assumed to be
$\hatrho_{\rm in}=|0_{\rm in}\rangle\langle 0_{\rm in}|$ throughout
this work.  The finite-temperature extension is rather
straightforward~\cite{Fukushima:2014sia}.  We note that this two-point
function (called the Wightman function) is nothing but
$\DF(t,\bp;t',-\bp)-\DR(t,\bp;t',-\bp)$ where
$\DF(t,\bp;t',\bp')$ and $\DR(t,\bp;t',\bp')$, respectively, represent
the time-ordered and the retarded propagators.  In the present work we
limit ourselves to the simplest case of complex scalar field theory
(i.e., scalar QED) under an external electric field, which is easily
translated to spinor matter.

\section{Stochastic quantization}

The key idea of the Stochastic Quantization is that one can quantize
field theories using a classical equation of motion with one
artificial axis (i.e.,\ quantum or Suzuki-Trotter
axis~\cite{Suzuki:1976}) denoted here by $\theta$ and with stochastic
variables $\eta(x,\theta)$.  We thus need to solve a
\textit{complex Langevin} equation, which turns out to be accompanied
by $\rmi$ in Minkowskian space-time.  Let us take a quick flash at the
way to retrieve free propagators.  As a matter of fact, a functional
formulation usually comes along with the time-ordered propagator,
whereas in the real-time problems we often need the retarded and
advanced propagators as well.  It is crucial, therefore, to establish
the correct description of them within the Stochastic Quantization
(without going back to the operator formalism).  For a free scalar
field theory the classical equation of motion reads,
\begin{equation}
 \frac{\partial\phi_p(t,\theta)}{\partial\theta}
  = \rmi\,\bigl[ -\partial_t^2 - \calE^2(\bp) \bigr]
    \phi_p(t,\theta) + \eta_p(t,\theta)
\label{eq:st_eq}
\end{equation}
with $\calE(\bp)\equiv\sqrt{\bp^2+m^2}$.  Here, we took the Fourier
transform with respect to spatial coordinates.  For our purpose to
cope with a time-dependent but spatially homogeneous background field,
it is convenient to keep $t$ not changed to the frequency.

In the complex scalar field theory of our interest, we need to
introduce another independent field $\tphi(t,\theta)$ and associated
stochastic variable $\teta_p(t,\theta)$.  In this partially Fourier
transformed representation we should define the average over the
stochastic variables as follows:
\begin{align}
 &\langle\eta_p(t,\theta)\,\teta_{p'}(t',\theta')\rangle_\eta
  = 2\,\delta(t\!-\!t') (2\pi)^3\delta^{(3)}(\bp\!+\!\bp')
    \delta(\theta\!-\!\theta'),\notag\\
 &\langle\eta_p(t,\theta)\,\eta_{p'}(t',\theta')\rangle_\eta
  = \langle\teta_p(t,\theta)\,\teta_{p'}(t',\theta')\rangle_\eta
  = 0 \;.
\label{eq:noise}
\end{align}
When we solve Eq.~\eqref{eq:st_eq}, the most useful boundary condition
is $\phi_p(t,0)=0$.  We could have taken a non-zero value, but then
we should supplement a proper subtraction in the end.  We can easily
find a formal solution of the complex Langevin equation given
explicitly as
\begin{equation}
 \phi_p(t,\theta) = \int_0^{\theta} \rmd\theta'\,
  \rme^{\rmi[-\partial_t^2 -\calE^2(\bp) + \rmi\epsilon] (\theta-\theta')}\,
  \eta_p(t,\theta') .
\label{eq:sol_phi}
\end{equation}
We inserted $\rmi\epsilon$ to guarantee the convergence in the
$\theta\to\infty$ limit, which corresponds to the $\rmi\epsilon$
prescription to derive the time-ordered propagator.

After taking the average we can simplify the expression of the
two-point function to reach the following form:
\begin{equation}
 \begin{split}
 & \langle\phi_p(t,\theta)\tphi_{p'}(t',\theta)\rangle_\eta
  = \frac{\rmi}{-\partial_t^2 - \calE^2(\bp) + \rmi\epsilon} \\
 &\quad\times \!\Bigl[ 1 \!-\!
  \rme^{2\rmi (-\partial_t^2-\calE_{\bp}^2+\rmi\epsilon)\theta} \Bigr]\,
  (2\pi)^3\delta^{(3)}(\bp+\bp')\, \delta(t-t') \;.
 \end{split}
\label{eq:freepropagator}
\end{equation}
When we take the $\theta\to\infty$ limit, the exponential oscillatory
term drops off, and the resultant expression is reduced to the
standard form of the time-ordered propagator,
i.e., $\DF(t,\bp;t',\bp')$.

It is a non-trivial question how to construct other types of the
propagators.  Since the creation and annihilation operators correspond
to the negative- and the positive-energy parts of the field operator,
it is then quite natural to decompose the stochastic variable as
$\eta_p(t,\theta)=\eta^{+}_p(t,\theta)+\eta^{-}_p(t,\theta)$
where
\begin{equation}
 \eta^{\pm}_p(t,\theta) \equiv \int_0^\infty \frac{\rmd\omega}{2\pi}
  \,\tilde{\eta}_p(\pm\omega,\theta)\, \rme^{\mp\rmi\omega t} \;.
\label{eq:de_eta}
\end{equation}
Here $\tilde{\eta}_p(\omega,\theta)$ represents the Fourier transform
of $\eta_p(t,\theta)$.  We also do the same for $\teta_p(t,\theta)$
and then $\delta(t-t')$ in Eq.~\eqref{eq:noise} is replaced with
$2\pi\delta(\omega+\omega')$ in the two-point function of
$\tilde{\eta}_p(\omega,\theta)$ and
$\tilde{\teta}_p(\omega',\theta')$.  Accordingly we can introduce
variants of Eq.~\eqref{eq:sol_phi}, namely:
\begin{equation}
 \phi_p^{\pm}(t,\theta) \equiv \int_0^\theta \rmd\theta'\,
  \rme^{\rmi[-\partial_t^2-\calE^2(\bp)+\rmi\epsilon] (\theta-\theta')}\,
  \eta_p^{\pm}(t,\theta') \;.
\label{eq:solution1}
\end{equation}
It is an important ingredient in our formulation to define:
\begin{equation}
 \psi_p^{\pm}(t,\theta) \equiv \int_0^\theta \rmd\theta'\,
  \rme^{-\rmi[-\partial_t^2-\calE^2(\bp)-\rmi\epsilon] (\theta-\theta')}\,
  \eta_p^{\pm}(t,\theta') \;,
\label{eq:solution2}
\end{equation}
which solves a slightly deformed equation of motion with the sign of
$\rmi$ flipped in Eq.~\eqref{eq:st_eq},  in other words, the equation
of motion derived from the sign-flipped action.  As we discuss later,
thus, $\psi_p^{\pm}(t,\theta)$ can be interpreted as the field along
the backward time path.

The time-ordered propagator involves only the components with
$\phi_p^{\pm}(t,\theta)$ and our main proposition here is to utilize
$\psi_p^{\pm}(t,\theta)$ as an additional building block of other
types of the propagators:
\begin{equation}
 \begin{split}
 & D_{\rm R}(t,\bp;t',\bp') \\
 &\qquad = \lim_{\theta\to\infty} \bigl\langle \phi^+_p\!(t,\theta)
  \tphi^-_{p'}\!(t',\theta) \!-\! \psi^-_p\!(t,\theta)
  \tpsi^+_{p'}\!(t',\theta) \bigr\rangle_\eta \;.
 \end{split}
\label{eq:retarded}
\end{equation}
We can also write the advanced propagator down in the same way by
means of an appropriate combination of $\phi_p^{\pm}(t,\theta)$ and
$\psi_p^{\pm}(t,\theta)$.  In view of Eq.~\eqref{eq:production},
therefore, we can identify an expression directly relevant to the
particle production as
\begin{equation}
 \begin{split}
 & D_{\rm F}(t,\bp;t',\bp') - D_{\rm R}(t,\bp;t',\bp') \\
 &\qquad = \lim_{\theta\to\infty} \bigl\langle \phi^-_p\!(t,\theta)
  \tphi^+_{p'}\!(t',\theta) \!+\! \psi^-_p\!(t,\theta)
  \tpsi^+_{p'}\!(t',\theta) \bigr\rangle_\eta \;.
 \end{split}
\label{eq:difference}
\end{equation}
We emphasize that, though our prescription may look ad-hoc at first
glance, this is a unique choice so that the convergence factor
$\rmi\epsilon$ has a right sign in the propagator as
$p_0^2-\calE^2(\bp)\pm\mathrm{sgn}(p_0)\,\rmi\epsilon$, after taking
the Fourier transform from $t$ to $p_0$.

\section{Time-dependent background field}

From now on we shall turn the time-dependent potential on, denoted by
$\calV_p(t)$, which yields a complex Langevin equation,
\begin{equation}
 \frac{\partial\phi^{\pm}_p(t,\theta)}{\partial\theta}
  = \rmi \bigl[ -\partial_t^2 + \calV_p(t) \bigr]
    \phi^{\pm}_p(t,\theta) + \eta^{\pm}_p(t,\theta)
\label{eq:st_eq2}
\end{equation}
and a similar one for $\psi_p^{\pm}(t,\theta)$ with $\rmi$ in the
right-hand side changed to $-\rmi$.  We assume a time-dependent but
spatially homogeneous electric field $\bE(t)$ and thus $\calV_p(t)$ is
given explicitly as
\begin{equation}
 \calV_p(t)=-m^2-[\bp-e\bA(t)]^2
\end{equation}
with $\bE(t)=-\partial_t \bA(t)$.  As long as $\calV_p(t)$ does not
involve momentum transfer, the spatial derivatives are diagonalized in
this partially Fourier transformed representation.  In the in- and the
out-states the interaction falls off, so that the asymptotic states
have $\calV_p(t\sim \ti)=-\calE^2_{\rm in}(\bp)$ and
$\calV_p(t\sim\tf)=-\calE^2_{\rm out}(\bp)$.  Let us demonstrate how
our formulas~\eqref{eq:production} and \eqref{eq:difference} work for
the estimate of the produced particle number.

We can easily solve \eqref{eq:st_eq2} for general $\calV_p(t)$ to find
the explicit form of the solution as
\begin{equation}
 \phi^{\pm}_p(t,\theta) = \int_0^\theta \rmd\theta'\,
  \rme^{\rmi[-\partial_t^2 + \calV_p(t)+\rmi\epsilon](\theta-\theta')}\,
  \eta^{\pm}_p(t,\theta')
\label{eq:sol_phi2}
\end{equation}
and we can solve for $\psi^{\pm}_p(t,\theta)$ as well.  We now get
ready to compute $D_{\rm R}(t,\bp;t',\bp')$ according to our
prescription.

The final answer should not depend on how we treat the $\eta$-average
as long as $\eta^{\pm}_p(t,\theta)$'s are generated consistently as
the Gaussian noise~\eqref{eq:noise}.  Instead of taking the Gaussian
average, we can simplify the calculation by means of
$\eta^{\pm}_p(t,\theta)$ decomposed with a complete set of the
solutions of the following equation of motion:
\begin{equation}
 \bigl[ -\partial_t^2 + \calV_p(t) \bigr] \chi^{\pm}_\omega(t)
  = \bigl[ \omega^2 - \calE_{\rm in}^2(\bp) \bigr] \chi^{\pm}_\omega(t) \;,
\label{eq:eom}
\end{equation}
where in the right-hand side, $\omega$ [or
$\omega^2-\calE_{\rm in}^2(\bp)$] is an eigenvalue to label the complete
set, and the superscript $\pm$ corresponds to the the boundary
condition,
\begin{equation}
 \chi^{\pm}_\omega(t\to \ti) \;\to\; \rme^{\mp\rmi\omega t} \;,
\label{eq:initial}
\end{equation}
which is chosen for convenience to meet the boundary condition of
Eq.~\eqref{eq:de_eta} at $t=\ti$.  Here, let us consider the electric
field along $x^3$ and take $\bA(t)=(0,0,A^3(t))$.  We note that
$\chi^{\pm}_\omega(t)$ correspond to the positive and negative energy
solutions of the classical equation of motion in
Ref.~\cite{Brezin:1970xf} and thus the Bogoliubov coefficients of
$\chi_\omega^\pm(t)$ yield the produced particle
spectrum~\cite{Brezin:1970xf,Nikishov:1970br}.

Because $\calV_p(t)$ is real,
$\chi_{-\omega}^{\mp}(t) = \chi_\omega^{\pm}(t)$ follows.  We can
deform the definition of positive- and negative-energy parts at
$t=\ti$ using this complete set:
\begin{equation}
 \eta^{\pm}_p(t,\theta) \equiv \int_0^\infty \frac{\rmd\omega}{2\pi}\,
  \tilde{\eta}_p(\pm\omega,\theta)\,\chi^{\pm}_\omega(t) \;,
\label{eq:de_eta2}
\end{equation}
which coincides with Eq.~\eqref{eq:de_eta} in the in-state at
$t=\ti$.  We would emphasize again that this parametrization is just
for practical convenience and we could have kept using the definition
of Eq.~\eqref{eq:de_eta} to come up to the same answer;  the
difference is whether we should cope with the complicated
$t$-dependent evolution operator in the exponential as seen in
Eq.~\eqref{eq:sol_phi2} or make it $t$-independent with the
complicated wave-function $\chi^{\pm}_p(t)$ (which is reminiscent of a
transition between the Schr\"{o}dinger and the Heisenberg pictures in
quantum mechanics).

With help of eigenfunctions of Eq.~\eqref{eq:eom} we can readily
derive the following form of the retarded propagator,
\begin{equation}
 \begin{split}
 &\DR(t,\bp;t',\bp') = (2\pi)^3 \delta^{(3)}(\bp+\bp') \\
 &\qquad\qquad\quad\times \int_{-\infty}^\infty\frac{\rmd\omega}{2\pi}\,
  \frac{\rmi\, \chi_\omega^{+}(t)\,\chi_\omega^{-}(t')}
  {\omega^2-\calE_{\rm in}^2(\bp)+\text{sgn}(\omega)\, \rmi\epsilon} \;.
 \end{split}
\end{equation}
For the particle production problem we need to calculate
$\DF-\DR$ which reads:
\begin{align}
 & \DF(t,\bp;t',\bp') - \DR(t,\bp;t',\bp')
  = (2\pi)^3\delta^{(3)}(\bp+\bp') \notag\\
 &\qquad \times \int_{-\infty}^0\frac{\rmd\omega}{2\pi}\,(-\rmi)2\pi
  \delta(\omega^2-\calE_{\rm in}(\bp)^2) \cdot
  \rmi\, \chi_\omega^{+}(t)\,\chi_\omega^{-}(t') \notag\\
 & = (2\pi)^3\delta^{(3)}(\bp+\bp')\,
  \frac{\chi^{+}_{-\calE_{\rm in}(\bp)}(t)\,\chi^{-}_{-\calE_{\rm in}(\bp)}(t')}
  {2\calE_{\rm in}(\bp)} \;.
\label{eq:dfdr}
\end{align}
We note that the delta function picks up an eigenvalue of
$\omega=-E_{\rm in}(\bp)$ only that makes the right-hand side of
Eq.~\eqref{eq:eom} vanishing!  Therefore,
$\chi^{\pm}_{-\calE_{\rm in}(\bp)}(t)$ satisfies the classical
equation of motion in the ordinary field theory.

With the initial condition~\eqref{eq:initial} the solution of the
equation of motion should behave like
$\chi^{-}_{-\calE_{\rm in}(\bp)}(t) = \rme^{-\rmi \calE_{\rm in}(\bp)t}$
near the in-state at $t=\ti$ and we can parametrize:
\begin{equation}
 \chi^{-}_{-\calE_{\rm in}(\bp)}(t) =
  \sqrt{\frac{\calE_{\rm in}(\bp)}{\calE_{\rm out}(\bp)}} \Bigl[
  \alpha_{\bp}\, \rme^{-\rmi \calE_{\rm out}(\bp)t} + \beta_{\bp}^\ast\,
  \rme^{\rmi \calE_{\rm out}(\bp)t} \Bigr] \;,
\end{equation}
near the out-state at $t=\tf$.  From these asymptotic forms it is easy
to find the following expression near the out-state as
\begin{align}
 & \DF(t,\bp;t',\bp') - \DR(t,\bp;t',\bp')
  = (2\pi)^3\delta^{(3)}(\bp+\bp') \notag\\
 & \times \frac{1}{2\calE_{\rm out}(\bp)} \biggl\{
  |\alpha_{\bp}|^2 \rme^{\rmi \calE_{\rm out}(\bp)(t-t')}
 +|\beta_{\bp}|^2 \rme^{-\rmi \calE_{\rm out}(\bp)(t-t')} \notag\\
 &\qquad\qquad\qquad\qquad + 2\text{Re}\bigl[\alpha_{\bp}\beta_{\bp}
  \rme^{-\rmi \calE_{\rm out}(\bp)(t+t')} \bigr] \biggr\} \;,
\end{align}
which recovers the results in Ref.~\cite{Fukushima:2014sia} and leads
to the well-known formula of the produced particle
spectrum~\cite{Brezin:1970xf,Nikishov:1970br}:
\begin{equation}
 \frac{\rmd N}{\rmd^3\bp} = \delta^{(3)}(0)\,|\beta_{\bp}|^2 \;.
\end{equation}
We make a remark that Eq.~\eqref{eq:eom} provides us with a basis of
the so-called over-the-barrier scattering picture for the Schwinger
mechanism~\cite{Brezin:1970xf,Nikishov:1970br} (see also
Refs.~\cite{Kluger:1991ib,Hebenstreit:2009km,Akkermans:2011yn} which
can be understood in this picture).

\section{Relation to other formalisms}

Now that we have reached the final expression of the particle
production, let us deepen a physical insight from the point of view of
both formal and numerical aspects.

As we already mentioned, $\psi_p^\pm(t,\theta)$ plays a similar role
to the field along the backward time path that appears in the
Schwinger-Keldysh or closed-time path (CTP)
formalism~\cite{Schwinger:1960qe,Keldysh:1964ud}.
In fact we can find a mapping to two-point functions in the canonical
quantization, that is:
\begin{align}
&\lim_{\theta\to\infty} \bigl\langle \phi^+_p(t,\theta)
 \tphi^-_{p'}(t',\theta) \bigr\rangle_\eta \!=\!
 \bigl\langle\Theta(t-t')\hphi_p(t,\theta)
 \hphi^\dag_{p'}(t',\theta)\bigr\rangle , \\
&\lim_{\theta\to\infty} \bigl\langle \phi^-_p(t,\theta)
 \tphi^+_{p'}(t',\theta) \bigr\rangle_\eta \!=\!
 \bigl\langle\Theta(t'-t)\hphi^\dag_{p'}(t',\theta)
 \hphi_p(t,\theta)\bigr\rangle , \\
&\lim_{\theta\to\infty} \bigl\langle \psi^+_p(t,\theta)
 \tpsi^-_{p'}(t',\theta) \bigr\rangle_\eta \!=\!
 \bigl\langle\Theta(t'-t)\hphi_p(t,\theta)
 \hphi^\dag_{p'}(t',\theta)\bigr\rangle , \\
&\lim_{\theta\to\infty} \bigl\langle \psi^-_p(t,\theta)
 \tpsi^+_{p'}(t',\theta) \bigr\rangle_\eta \!=\!
 \bigl\langle\Theta(t-t')\hphi^\dag_{p'}(t',\theta)
 \hphi_p(t,\theta)\bigr\rangle
\end{align}
with $\Theta(t)$ being the Heaviside step function.  We use the hat to
indicate the quantum operator.  The Schwinger-Keldysh formalism
consists of $2\times 2$ matrix propagators which we can construct from
the above two-point functions as
\begin{align}
 \begin{split}
 & D_{++}(t,\bp;t',\bp') \equiv \bigl\langle \T\bigl[\hphi_p(t,\theta)
  \hphi^\dag_{p'}(t',\theta)\bigr]\bigr\rangle \\
 &\quad = \lim_{\theta\to\infty} \bigl\langle \phi^+_p(t,\theta)
  \tphi^-_{p'}(t',\theta) + \phi^-_p(t,\theta)
  \tphi^+_{p'}(t',\theta) \bigr\rangle_\eta \;,
 \end{split}
\label{eq:timeordered}
\\
 \begin{split}
 & D_{--}(t,\bp;t',\bp') \equiv \bigl\langle \tT\bigl[\hphi_p(t,\theta)
  \hphi^\dag_{p'}(t',\theta)\bigr]\bigr\rangle \\
 &\quad = \lim_{\theta\to\infty} \bigl\langle \psi^+_p\!(t,\theta)
  \tpsi^-_{p'}(t',\theta) + \psi^-_p(t,\theta)
  \tpsi^+_{p'}(t',\theta) \bigr\rangle_\eta \;,
 \end{split}
\label{eq:reversetimeordered}
\\
 \begin{split}
 & D_{+-}(t,\bp;t',\bp') \equiv \bigl\langle \hphi^\dag_{p'}(t',\theta)
  \hphi_p(t,\theta)\bigr\rangle \\
 &\quad = \lim_{\theta\to\infty} \bigl\langle \phi^-_p(t,\theta)
  \tphi^+_{p'}(t',\theta) + \psi^-_p(t,\theta)
  \tpsi^+_{p'}(t',\theta) \bigr\rangle_\eta \;,
 \end{split}
\label{eq:cross1}
\\
 \begin{split}
 & D_{-+}(t,\bp;t',\bp')\equiv \bigl\langle \hphi_p(t,\theta)
  \hphi^\dag_{p'}(t',\theta)\bigr\rangle \\
 &\quad = \lim_{\theta\to\infty} \bigl\langle \phi^+_p\!(t,\theta)
  \tphi^-_{p'}(t',\theta) + \psi^+_p(t,\theta)
  \tpsi^-_{p'}(t',\theta) \bigr\rangle_\eta
 \end{split}
\label{eq:cross2}
\end{align}
where $\T$ and $\tT$, respectively, denote the time and reversed-time
ordered products.  By using the explicit solutions
\eqref{eq:solution1} and \eqref{eq:solution2}, we can show that these
propagators are equivalent to those defined in
Ref.~\cite{Epelbaum:2014yja}.  Thus, we can regard
$\psi_p^{\pm}(t,\theta)$ as the positive and negative energy fields
along the backward time path and our formulation encompasses the
precise structure of the perturbation theory in the Schwinger-Keldysh
formalism.

\begin{figure}
 \includegraphics[width=0.9\columnwidth]{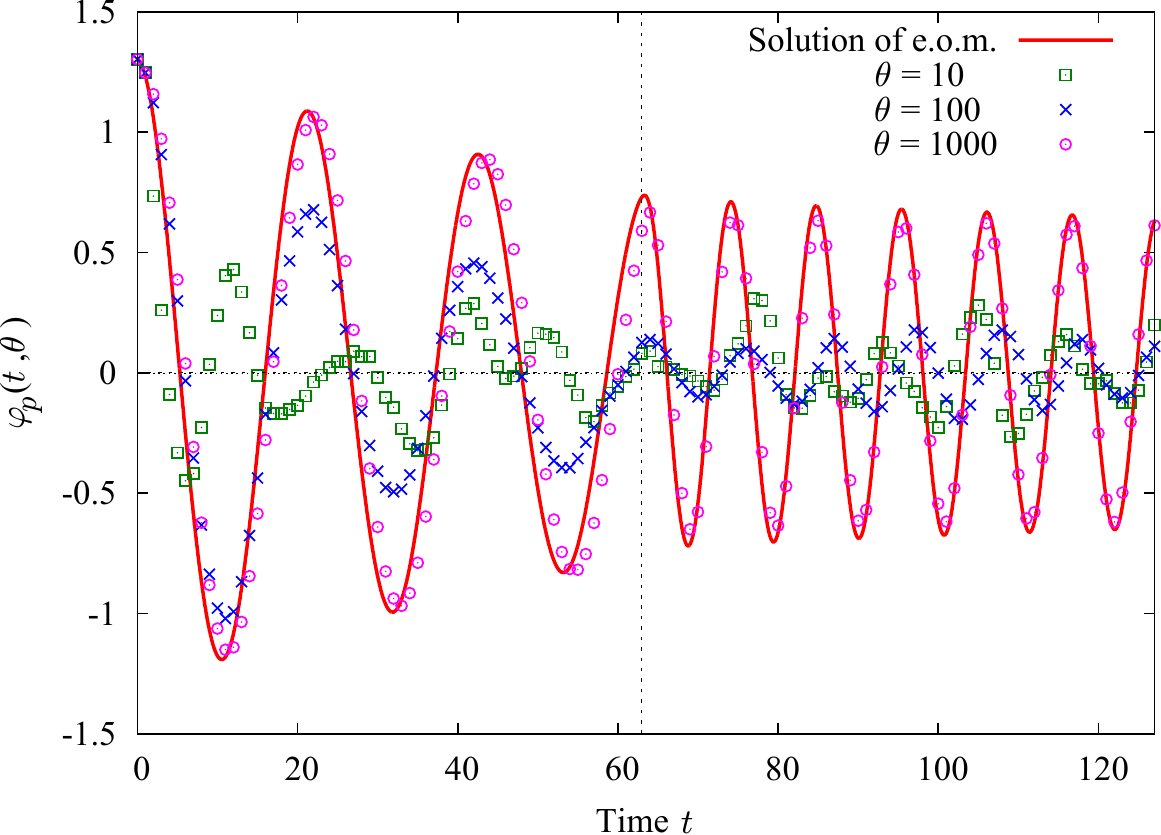}
 \caption{Evolution of the averaged field variable
   $\varphi_p(t,\theta)$ from $\ti=0$ with increasing $\theta$.  A
   pulse electric field is imposed around $t=t_0$.  The boundary
   condition at $t=\ti$ is specified as an outgoing form:
   $\varphi_p(t,\theta)\propto
   \rme^{\rmi\calE_{\rm in}\ti}$.}
 \label{fig:freeprop}
\end{figure}

For the rest of this paper, we will address the relation to the
classical statistical simulation~\cite{Gelis:2013oca}.  Let us
consider a numerical simulation with fixed values of
$\phi(\ti,\theta)$ and $\dot{\phi}(\ti,\theta)$ [or
$\phi(\ti+\Dt,\theta)$] to solve Eq.~\eqref{eq:st_eq}.  We then
perform the $\eta$-average except at $t=\ti$ and $\ti+\Dt$.  Taking
the $\theta$-average can significantly stabilize the
$\theta$-oscillation and reduce the computational cost.  More
specifically, the $\theta$-averaged field as defined by
\begin{equation}
 \varphi_p(t,\theta) \equiv
  \theta^{-1}\int_0^\theta \rmd\theta'\,\phi_p(t,\theta')\;,
\end{equation}
approaches the solution of the equation of motion~\eqref{eq:eom}.  We
can clearly confirm it in Fig.~\ref{fig:freeprop} in the presence of
an electric field pulsed around $t=t_0$, which is chosen specifically as
\begin{equation}
 \bA(t)= \left(0,0, \frac{E_0}{\mathfrak{w}}
  \bigl[\tanh\mathfrak{w}(t-t_0)+1\bigr]\right) \;.
\end{equation}
Physical quantities are all made dimensionless by the time step $\Dt$
and the site number along the $t$-axis is chosen as $N_t=256$.  The
$\theta$-axis is discretized with $\Delta\theta=5\times 10^{-3}$
(which means that we update the $\theta$-evolution $2\times 10^5$
times to get the results at $\theta=1000$).  We choose $p_3=0$ and
$\calE_{\rm in}(\bp) = \sqrt{(p_1)^2+(p_2)^2+m^2}=12\times
(2\pi/N_t)$, so that there are 12 periods included along the
$t$-direction from $t=0$ to $(N_t-1)\Dt$ if not affected by the
electric field.  We postulate a short life time for the electric
field:  $\mathfrak{w}=5\calE_{\rm in}(\bp)$ for a fixed momentum $\bp$
and the it stands at $t_0=63\Dt$ (i.e., a quarter of the whole time
range).

To manifest the effect of the electric field, we specifically adopt:
$|e|E_0/\mathfrak{w}=(\sqrt{3}/2)\calE_{\rm in}(\bp)$, and then
$\calE_{\rm out}(\bp)=2\calE_{\rm in}(\bp)$.  With this choice we see
that the results in Fig.~\ref{fig:freeprop} is quite reasonable;
there are 3 and 6 periods of the oscillation from $t=0$ to $t_0$ and
from $t=t_0$ to $2t_0$, respectively, observed in
Fig.~\ref{fig:freeprop}.  We note that $\epsilon=5\times 10^{-3}$ is
used for numerical stability.  On the technical level it is the most
tough part to avoid unphysical ``run-away'' flows in $\theta$, which
is overcome here by implementing the Crank-Nicolson
method~\cite{Anzaki:2014hba}.

We imposed an outgoing initial condition as
$\varphi_p(\ti,\theta)=(1/\sqrt{2\calE_{\rm in}(\bp)})
\rme^{-\rmi \calE_{\rm in}(\bp)\ti}$ at $t=\ti$ and $t=\ti+\Dt$ in our
Stochastic Quantization simulation, which is the right choice to
evaluate the production rate in the ordinary
procedure~\cite{Brezin:1970xf,Nikishov:1970br}.  Also, we numerically
solved the equation of motion \eqref{eq:eom} in the presence of $\bA(t)$ with the
same initial condition as shown by a solid curve in
Fig.~\ref{fig:freeprop}.  It is clear that the Stochastic Quantization
output converges to the solution of the equation of motion as it
should.  It should be mentioned that the decomposition to positive-
and negative-energy parts with $\eta^{\pm}_p(t,\theta)$ is now
effectively taken into account in our procedure to impose the outgoing
initial condition.  Since the convergence to the solution of the
equation of motion guarantees that we can reproduce correct
$\rmd N/\rmd^3\bp$, we would not explicitly evaluate it.

Let us comment on the relation to the classical statistical
simulation~\cite{Berges:2007ym,Fukushima:2006ax,Gelis:2013rba} here.
If we compute $\langle \phi_p(t) \rangle$, as seen in
Fig.~\ref{fig:freeprop}, the Stochastic Quantization leads to the
solution of the equation of motion.  More generally, if we are allowed
to make an approximation for an operator $\mathcal{O}[\phi]$ that
$\langle\mathcal{O}[\phi]\rangle_t
\approx \mathcal{O}[\langle\phi\rangle_t]$ for a given initial
condition, this is nothing but the calculation procedure in the
classical statistical simulation.  The initial state should
accommodate quantum fluctuations described by the initial Wigner
function, and so we should perform the ensemble average with
fluctuating initial conditions in general.  (For the present purpose
to investigate the vacuum physics the $\rmi\epsilon$ prescription is
sufficient.)  We would emphasize that such a derivation of the
classical statistical simulation sheds light on the structure of the
approximation, e.g., the renormalization problem as addressed in
Ref.~\cite{Epelbaum:2014yja}.


\section{Summary}

In summary, in this work, we gave a derivation of the standard formula
for the Schwinger mechanism with Stochastic Quantization.  The most
non-trivial part was how to prescribe the retarded propagator, in such
a way that the $\theta$-integration is properly regulated.  We
decomposed the stochastic variables into positive- and negative-energy
parts, and this corresponds to imposing a proper initial condition in
the numerical simulation.  We showed that our machinery has a natural
connection to the closed-time path formalism and we presented our
numerical results that converge to the correct answer.

Our formulation on the basis of Stochastic Quantization has potential
applications to variety of real-time physics problems.  Apart from the
particle production issue, one of the most interesting extensions
would be the computation of the spectral functions and the transport
coefficients.  We are now making a progress in this direction.

\section*{Acknowledgments}
We thank Ryoji~Anzaki, Yoshimasa~Hidaka, Takashi~Oka, Shoichi~Sasaki
for fruitful discussions.
K.~F.\ was supported by JSPS KAKENHI Grant Number 24740169.
T.~H.\ was supported by JSPS Research Fellowships for Young
Scientists.


\end{document}